\documentclass[11pt]{article}

\usepackage{epsfig,subfigure}
\usepackage{amssymb}
\usepackage[fleqn]{amsmath}
\usepackage{color}
\usepackage{arydshln}

\newcommand{\define}{\stackrel{\triangle}{=}}
\def\QED{\mbox{\rule[0pt]{1.5ex}{1.5ex}}}

\def\new{\mbox{\texttt{new}}}

\usepackage{graphicx}
\usepackage{color}
\usepackage[dvipsnames]{xcolor}

\definecolor{armygreen}{rgb}{0.29, 0.33, 0.13}

\usepackage{amssymb}
\usepackage{amsmath,amsfonts,amssymb}
\usepackage{verbatim}
\usepackage{stfloats}
\usepackage[bookmarks=false]{}
\usepackage{algorithm}
\usepackage{algcompatible}

\newtheorem{theorem}{Theorem}
\newtheorem{corollary}{Corollary}

\newtheorem{lemma}{Lemma}

\newcommand\m[1]{%
\mbox{\small #1}%
}

\newcommand\blfootnote[1]{%
  \begingroup
  \renewcommand\thefootnote{}\footnote{#1}%
  \addtocounter{footnote}{-1}%
  \endgroup
}

\begin{document}
\date{}
\title{
Optimal Download Cost of Private Information Retrieval\\
for Arbitrary Message Length
}
\author{ \normalsize Hua Sun and Syed A. Jafar \\
}

\maketitle

\blfootnote{Hua Sun (email: huas2@uci.edu) and Syed A. Jafar (email: syed@uci.edu) are with the Center of Pervasive Communications and Computing (CPCC) in the Department of Electrical Engineering and Computer Science (EECS) at the University of California Irvine. 
}

\begin{abstract}
A private information retrieval scheme is a mechanism that allows a user to retrieve any one out of $K$ messages from $N$ non-communicating replicated databases, each of which stores all $K$ messages, without revealing anything about the identity of the desired message index to any individual database. If the size of each message is $L$ bits and the  total download required by a PIR scheme from all $N$ databases  is $D$ bits, then $D$ is called the download cost and the ratio $L/D$ is called an achievable rate.  For fixed $K,N\in\mathbb{N}$, the capacity of PIR, denoted by $C$, is the supremum of  achievable rates over all PIR schemes and over all message sizes, and was recently shown to be $C=(1+1/N+1/N^2+\cdots+1/N^{K-1})^{-1}$. In this work, for arbitrary $K, N$, we explore the minimum download cost $D_L$ across all PIR schemes (not restricted to linear schemes) for arbitrary message lengths $L$ under arbitrary choices of alphabet (not restricted to finite fields) for the message and download  symbols. If the same $M$-ary alphabet is used for the message and download symbols, then we show that the optimal download cost in $M$-ary symbols is $D_L=\lceil\frac{L}{C}\rceil$.  If the message symbols are in $M$-ary alphabet and the downloaded symbols are in $M'$-ary alphabet, then we show that the optimal download cost in $M'$-ary symbols,  $D_L\in\left\{\left\lceil \frac{L'}{C}\right\rceil,\left\lceil \frac{L'}{C}\right\rceil-1,\left\lceil \frac{L'}{C}\right\rceil-2\right\}$, where $L'= \lceil L \log_{M'} M\rceil$.
\end{abstract}

\allowdisplaybreaks
\section{Introduction}
In the private information retrieval (PIR) problem \cite{PIRfirst, PIRfirstjournal}, we have $K$ messages, stored at $N$ distributed and non-communicating databases.  A PIR scheme allows a user  to retrieve any one of the $K$ messages, while revealing no information to any individual database about the retrieved message index. Typical quality measures of PIR schemes include communication complexity \cite{PIRfirst, PIRfirstjournal, Ambainis, Beimel_Ishai_Kushilevitz, Beimel_Ishai_Kushilevitz_Raymond, YekhaninPhd, 2PIR}, computational overhead \cite{Beimel_Ishai_Malkin, Gertner_Goldwasser_Malkin, Crescenzo_Ishai_Ostrovsky}, storage overhead \cite{Shah_Rashmi_Kannan, Chan_Ho_Yamamoto, Fazeli_Vardy_Yaakobi, Tajeddine_Rouayheb, Rao_Vardy, Blackburn_Etzion, Blackburn_Etzion_Paterson, Banawan_Ulukus, Zhang_Wang_Wei_Ge}, upload cost \cite{Sun_Jafar_PIR},  download cost \cite{Shah_Rashmi_Kannan, Chan_Ho_Yamamoto, Tajeddine_Rouayheb, Sun_Jafar_PIR, Banawan_Ulukus}, and rate \cite{Sun_Jafar_PIR, Tajeddine_Rouayheb, Banawan_Ulukus}. In this work we will focus on download cost and rate. If the size of each message is $L$ bits and the  total download required by a PIR scheme from all $N$ databases  is $D$ bits, then $D$ is called the download cost and the ratio $L/D$ is called an achievable rate.  The capacity of PIR,  denoted by $C$, is defined to be the supremum of  achievable rates over all PIR schemes and over all message sizes.  It was shown recently in  \cite{Sun_Jafar_PIR} that\footnote{We will use the symbol $C$ to represent the expression in (\ref{eq:C}) throughout this paper.}
\begin{eqnarray}
C = \left(1 + \frac{1}{N} + \frac{1}{N^2} + \cdots + \frac{1}{N^{K-1}}\right)^{-1} \label{eq:C}
\end{eqnarray}
The reciprocal of capacity, $1/C$, similarly represents the infimum of download cost per message bit over all PIR schemes and over all message sizes. Fundamental information theoretic measures such as these are essentially asymptotic in character, involving  limits as message lengths $L\rightarrow \infty$. Remarkably, \cite{Sun_Jafar_PIR} shows that these asymptotically optimal values are also achieved exactly when the message length parameter $L$ is any integer multiple of $N^K$. However, since in practice the message length parameter $L$ can be arbitrary, an important question that remains open is to determine optimal download cost and rate values for arbitrary fixed values of $L$, in particular when $L$ is not an integer multiple of $N^K$. In this work, we explore the minimum download cost across all PIR schemes (not restricted to linear schemes) for arbitrary message lengths under arbitrary choices of alphabet (not restricted to finite fields) for the message and download symbols. If the same $M$-ary alphabet is used for the message and download symbols, then we show that the optimal download cost in $M$-ary symbols is $D_L=\lceil\frac{L}{C}\rceil$.  If the message symbols are in $M$-ary alphabet and the downloaded symbols are in $M'$-ary alphabet, then we show that the optimal download cost in $M'$-ary symbols,  $D_L\in\left\{\left\lceil \frac{L'}{C}\right\rceil,\left\lceil \frac{L'}{C}\right\rceil-1,\left\lceil \frac{L'}{C}\right\rceil-2\right\}$, where $L'= \lceil L \log_{M'} M\rceil$. Correspondingly, the maximum achievable rate  is automatically characterized in every case as $L/D_L$.

{\it Notation: $\mathbb{N}$ is the set of natural numbers. For  integers $Z_1, Z_2, Z_1 \leq Z_2$, we use the compact notation $[Z_1:Z_2]=\{Z_1, Z_1+ 1,\cdots, Z_2\}$. Similarly, $A_{[Z_1:Z_2]} \define \{A_{Z_1},A_{Z_1 + 1},\cdots,A_{Z_2}\}$ for any variable $A$.  The notation $X \sim Y$ is used to indicate that $X$ and $Y$ are identically distributed. The notation $|A|$ is used to denote the cardinality of a set when $A$ is a set, and the length of a tuple when $A$ is a tuple.  For sets $S_1, S_2$, we define $S_1/S_2$ as the set  of elements that are in $S_1$ and not in $S_2$. For a permutation function $\lambda(\cdot)$ applied to some $l$-tuple $U=(U(1),U(2),\cdots,U(l))$, we will allow some abuse of notation to write $\lambda(U)=(U(\lambda(1)),U(\lambda(2)),\cdots,U(\lambda(l)))$. }

\section{Problem Statement}\label{sec:model}

There are $K$ messages $W_1, \cdots, W_K$, each of which is an arbitrary string  of length $L$ comprised of $M$-ary symbols. 
\begin{eqnarray}
W_k = \Big( W_k(1), W_k(2), \cdots, W_k(L) \Big) \in [0: M-1]^L &&\forall k\in[1:K]
\end{eqnarray}
Note that there are $M^L$ possible distinct realizations of each message.



\noindent There are $N$ databases. Each database stores all the messages $W_1, \cdots, W_K$. 

Depending upon the desired message index $\theta\in[1:K]$, the user follows one of $K$ strategies. These strategies are specified in terms of $KN$ random queries, $Q_n^{[\theta]}$, $\forall n\in[1:N], \forall \theta\in[1:K]$ that are privately generated by the user a-priori, i.e., without any knowledge of the message realizations. In order to retrieve $W_\theta$,   the user sends the query $Q_n^{[\theta]}$ to the $n$-th database, $\forall n\in[1:N]$.

Upon receiving $Q_n^{[\theta]}$, the $n$-th database returns an answering string $A_n^{[\theta]}$, which is a  function of $Q_n^{[\theta]}$ and the data stored (i.e., messages $W_1, \cdots, W_K$). The answering string $A_n^{[\theta]}$ is   comprised of $M'$-ary symbols, $A_n^{[\theta]}  \in [0:M'-1]^{|A_n^{[\theta]}|}$. 

From all the information that is now available to the user ($A_{[1:N]}^{[\theta]},  Q_{[1:N]}^{[\theta]}$), he must be able to correctly decode the desired message $W_\theta$. That is, the following correctness constraint must be satisfied.
\begin{eqnarray}
[\mbox{Correctness}] ~~W_\theta ~\mbox{is a  function of} ~ A_{[1:N]}^{[\theta]},  Q_{[1:N]}^{[\theta]}. \label{corr}
\end{eqnarray}

To protect the user's privacy, the query presented to each database must be identically distributed regardless of the desired message index. 
\begin{eqnarray}
[\mbox{Privacy}] ~~Q_n^{[\theta]}&\sim&Q_n^{[\theta']}, ~~~\forall   \theta,\theta'\in[1:K], n\in[1:N]. \label{privacy}
\end{eqnarray}



\noindent The download cost, $D$, for a PIR scheme is the maximum value (across all random realizations of queries) of the total number of $M'$-ary symbols downloaded by the user from all the databases. 
\begin{eqnarray}
D&=&\max\sum_{n=1}^N |A_n^{[\theta]}| 
\end{eqnarray}
Our goal is to characterize the optimal (minimum over all PIR schemes) download cost $D_L$, for \emph{arbitrary} fixed message size $L$. The optimality is across \emph{all} PIR schemes, i.e., including non-linear PIR schemes.

\section{Results}\label{sec:main}
\subsection{Optimal Download Cost for Matching Alphabet $(M=M')$}\label{sec:match}
Consider the setting where the messages and downloads are comprised of symbols from the same alphabet, i.e., $M=M' \in\mathbb{N}/\{1\}$. Our main result for this setting appears in the following theorem.
\begin{theorem} For PIR with  $N\in\mathbb{N}$ databases, each storing all $K\in\mathbb{N}$ messages, each message comprised of $L\in\mathbb{N}$ symbols from $M$-ary alphabet, $M\in\mathbb{N}/\{1\}$, where the downloads are  comprised of symbols from the same $M$-ary alphabet, the optimal download cost is $D_L=\left\lceil\frac{L}{C}\right\rceil$ $M$-ary symbols. 
\label{thm:download}
\end{theorem}
The proof of converse (i.e., the impossibility claim) of Theorem \ref{thm:download} follows from the capacity result of \cite{Sun_Jafar_PIR} and appears in Section \ref{proof:converse}. The achievability is proved, first for the case $L=N^{K-1}$ in Section \ref{proof:achLmin}, and then for arbitrary $L$ in Section \ref{proof:achLgen}. 

Based on Theorem \ref{thm:download}, the following observations are in order.
\begin{enumerate}
\item Given the message size and alphabet constraints, since the minimum download cost corresponds to the maximum rate, Theorem \ref{thm:download} equivalently characterizes the optimal rate for arbitrary message size in the matching alphabet case, as $L/\lceil \frac{L}{C} \rceil$.
\item Reference \cite{Shah_Rashmi_Kannan} shows that when $K \geq 2$ and  $N \geq L + 1$, then the optimal download is $D_L = L+1$. This result can  be recovered as a special case of Theorem \ref{thm:download} by noting that when $K \geq 2$ and $N \geq L+1$,
\begin{eqnarray}
D_L&=&\left\lceil\frac{L}{C}\right\rceil\\
&=&L + \left\lceil L\left( \frac{1}{N} + \frac{1}{{N^2}} + \cdots + \frac{1}{{N^{K-1}}}\right) \right\rceil \\
&=&L+1\label{eq:leap}
\end{eqnarray}
where (\ref{eq:leap}) follows because $0< \frac{1}{N} + \frac{1}{{N^2}} + \cdots + \frac{1}{{N^{K-1}}} < \frac{1}{N-1} \leq \frac{1}{L}$.
Theorem \ref{thm:download} completes the picture by characterizing the optimal download cost for all $N, K, L$.

\item Reference \cite{Sun_Jafar_PIR} presents a PIR scheme which achieves a rate equal to capacity $C$ if $L=nN^K$ where $n\in\mathbb{N}$ is any positive integer, so that the corresponding download is $D=\frac{L}{C}$. This result can be recovered as a special case of Theorem \ref{thm:download} by noting that when $L = n N^K$, then $\frac{L}{C} = n N^K (1+1/N +\cdots + 1/N^{K-1}) = n N(1 + N + \cdots + N^{K-1})$ is a positive integer so that $D_L  = \left\lceil\frac{L}{C}\right\rceil= \frac{L}{C}$.

\item A naive extension of the PIR scheme of \cite{Sun_Jafar_PIR}  to the setting when $L$ is not an integer multiple of $N^K$, is obtained by padding zeros to each message so that the message lengths are rounded up to the closest integer multiple of $N^K$.
 The gap between the download cost of the naive scheme and the optimal download cost in Theorem \ref{thm:download} can be unbounded. For an example, if $L = N^{K-1}$, then the download cost of the naive scheme is $D =  N^K/C$, while the optimal download cost is $D_L = \left\lceil\frac{L}{C}\right\rceil = N^{K-1}/C$.
 
 \item In the absence of any constraints on message lengths, we know from \cite{Sun_Jafar_PIR} that the maximum achievable rate across all PIR schemes is the capacity $C$. For constrained message length $L$, Theorem \ref{thm:download} shows that the maximum achievable rate 
 is $L/D_L = L/\left\lceil\frac{L}{C}\right\rceil$ which is in general less than $C$. The message length $L=N^{K-1}$ is particularly significant in light of Theorem \ref{thm:download}, because this is the shortest message length for which the achieved rate equals the capacity $C$. This is seen as follows. In order to achieve the capacity, the download cost must be $D=\frac{L}{C}=D_L$ which must be a positive integer value. But if $L<N^{K-1}$, then
\begin{eqnarray}
D = \frac{L}{C}= L \left(1 + \frac{1}{N} + \cdots + \frac{1}{N^{K-1}}\right)= L \left(\frac{1 + N + \cdots + N^{K-1}}{N^{K-1}}\right) \notin \mathbb{N}
\end{eqnarray}
because $N^{K-1}$ and $1+N+\cdots+N^{K-1}$ are co-prime. This is verified, e.g., through Bezout's identity,
\begin{eqnarray}
N^{K-1}(N)+(1+N+\cdots+N^{K-1})(1-N)&=&1
\end{eqnarray}
\end{enumerate}

\subsection{Optimal Download Cost for Mismatched Alphabet $(M\neq M')$}
Now consider PIR schemes with mismatched alphabet, i.e., the messages are represented in $M$-ary alphabet, and the downloaded symbols are in $M'$-ary alphabet, $M'\neq M$. For this setting the optimal download cost to within 2 symbols is characterized in the following theorem.
 
\begin{theorem}\label{thm:gen} 

For PIR with  $N\in\mathbb{N}$ databases, each storing all $K\in\mathbb{N}$ messages, each message comprised of $L\in\mathbb{N}$ symbols from $M$-ary alphabet, $M\in\mathbb{N}/\{1\}$, where the downloads are  comprised of symbols from  $M'$-ary alphabet, $M'\in\mathbb{N}/\{1\}$, $M'\neq M$, the optimal download cost  $D_L\in\left\{\left\lceil \frac{L'}{C}\right\rceil,\left\lceil \frac{L'}{C}\right\rceil-1,\left\lceil \frac{L'}{C}\right\rceil-2\right\}$,  where $L'= \lceil L \log_{M'} M\rceil$.
\end{theorem}
The proof of Theorem \ref{thm:gen} appears in Section \ref{proof:thmgen}. 

The following observations place Theorem \ref{thm:gen}  in perspective.
\begin{enumerate}
\item The proof of Theorem \ref{thm:gen} presented in Section \ref{proof:thmgen} shows that the download cost $\left\lceil \frac{ \lceil L \log_{M'} M\rceil }{C}\right\rceil$  is always achievable, and the download cost for any PIR scheme cannot be less than $\lceil \frac{L \log_{M'} M}{C} \rceil$. Therefore, in particular, when $\left\lceil \frac{ \lceil L \log_{M'} M\rceil }{C}\right\rceil = \lceil \frac{L \log_{M'} M}{C} \rceil$,  the \emph{exact} optimal download cost is $D_L = \lceil \frac{L \log_{M'} M}{C} \rceil$.
\item It is easy to create examples where mismatched alphabet leads to less efficient PIR schemes than possible with matched alphabets. However, this is not always the case. The following examples show  how mismatched alphabet can in some cases be beneficial in terms of rate relative to matched alphabet. Consider $N=2, K=2$,  $L=3,  M=9$. Here $ C=2/3$. The highest rate achievable with matched alphabet $(M'=M)$ is $\frac{L}{\lceil L/C\rceil}=3/5<C$ whereas the rate achieved with the mismatched alphabet $M'=3<M$, is  $\frac{L\log_{M'}M}{\lceil L'/C\rceil}=2/3=C$. Similarly one can construct examples with $M'>M$ where mismatched alphabet produces a higher rate than the best possible with matched alphabet, e.g., $N=2,K=2, L=3, M=4$ where the best rate with matched alphabet is again $3/5<C$, but the mismatched alphabet $M'=8$ achieves rate $2/3=C$.
\end{enumerate}
\section{Proof of Theorem \ref{thm:download}: Converse}\label{proof:converse}
The converse for Theorem \ref{thm:download} is the impossibility claim, i.e., that no PIR scheme with matched alphabet $(M=M')$ can achieve a download cost smaller than $D_L=\lceil\frac{L}{C}\rceil$. This is proved as follows.

The message realizations are arbitrary, as is the choice of the desired message index $\theta\in[1:K]$. By arbitrary, what is meant is that all realizations are possible. Therefore the PIR scheme must work for every possible realization of message symbols and $\theta$. Any PIR scheme that works for arbitrary realizations, will also work if they are uniformly randomly generated. Therefore, for the converse argument let us assume  uniform distributions on  the realizations of message symbols, and on $\theta$. The advantage of assigning a distribution to these arbitrary quantities is that we are able to use the information theoretic formulation of the PIR problem as in \cite{Sun_Jafar_PIR}, and the upper bounds on rate that are derived in \cite{Sun_Jafar_PIR} are also applicable in our current setting. In particular, $C$ is still an upper bound on the achievable rate of a PIR scheme with arbitrary message realizations and $\theta$ and arbitrary message length $L$. Since capacity is an upper bound on the rate of all PIR schemes, $C\geq L/D_L$, so that $D_L\geq \frac{L}{C}$, and because $D_L\in\mathbb{N}$, we must have $D_L\geq \left\lceil\frac{L}{C}\right\rceil$. 

\section{Proof of Theorem \ref{thm:download}: Achievability for $L=N^{K-1}$}\label{proof:achLmin}
In \cite{Sun_Jafar_PIR}, it is shown that the capacity (and the corresponding optimal download cost) of PIR is achievable when $L = N^K$ bits.  Here we present a more efficient PIR scheme to show that a smaller message size, $L = N^{K-1}$ bits ($M$-ary symbols in general), is  sufficient to achieve a rate equal to $C$ (and the optimal download cost)  when the alphabets are matched, i.e., $M=M'$. This PIR scheme is significant because (as noted in Section \ref{sec:match}) $L=N^{K-1}$ is  the smallest message size needed to achieve capacity, and also because it is the key ingredient that will allow us to subsequently expand the achievability proof to arbitrary $L$ in Section \ref{proof:achLgen}. Note that since the $N=1$ case is trivial (optimal to download all messages), we will consider only $N\geq 2$ in this section.

The PIR scheme that we present here is closely related to the capacity achieving PIR scheme  presented in \cite{Sun_Jafar_PIR}. For both schemes the queries are comprised only of sums of symbols from various messages. Since our new scheme considers $M$-ary alphabet, the ``sums" are interpreted as modulo-$M$ sums. In both schemes no symbol appears more than once in the query for any particular database. The difference between the two schemes lies in the requirement of symmetry across databases. Recall that the PIR scheme of \cite{Sun_Jafar_PIR} is based on the iterative application of three steps corresponding to symmetry across databases, symmetry across messages within the query to each database, and exploiting side information. The key to reducing the message size from $L = N^{K}$ to $L = N^{K-1}$ is to eliminate the requirement of symmetry across databases. Therefore, the new PIR scheme for $L=N^{K-1}$, formalized in the {\bf Q-Gen} Algorithm in Section \ref{sec:Q-Gen}, is based on the iterative application of the following two steps.
\begin{enumerate}
\item[(1)] {\it Enforcing Message Symmetry within the Queries to Each Database}: The goal is to make the queries to a database symmetric with respect to messages. For instance if the query to database $1$ includes $l$ instances of sums of  symbols from messages $W_1,W_2,W_3$, then it must  include $l$ instances of sums of symbols from each of the $\binom{K}{3}$ combinations of $3$ messages.  Message symmetry is defined formally in Section \ref{sec:def}.  The procedure is formalized in the {\bf M-Sym} Algorithm, presented in Section \ref{sec:twosub}. All the queries that do not involve desired message symbols ($\mathcal{I}$ terms in the {\bf Q-Gen} Algorithm) are introduced only through the {\bf M-Sym} algorithm. 

\item[(2)] {\it Exploiting Side Information}: The goal of this step is to exploit queries from other databases that were added to enforce message symmetry (and do not contain desired message symbols), as side information to construct new queries which are sums of symbols from desired message and the side information available from other databases. This step is formalized in the {\bf Exploit-SI} Algorithm, presented in Section \ref{sec:twosub}. Except for an initialization step, all the queries involving desired message symbols ($\mathcal{M}$ terms in the {\bf Q-Gen} Algorithm) are introduced only through the {\bf Exploit-SI} algorithm. 
\end{enumerate}
Let us start with a few simple examples for small $K,N$ values to illustrate the key ideas.

\subsection{Examples}\label{sec:ex}
\subsubsection*{$K=2$ Messages, $N=2$ Databases, $L=N^{K-1}=2$ Symbols Per Message}\label{sec:k2n2}
Let $[a_1,a_2]$ represent a random permutation of $L=2$ symbols from $W_1$. Similarly, let $[b_1, b_2]$ represent an independent random permutation of $L=2$ symbols from $W_2$. The key to the privacy of the scheme is that these random permutations are generated privately by the user and are unknown to the databases. 

Suppose the desired message is $W_1$, i.e., $\theta = 1$.
The PIR scheme always starts by requesting the first desired symbol (in this case, $a_1$) from the first database (DB1). Applying Step (1), we achieve message symmetry by including  $b_1$ from DB1. Next we apply Step (2) to exploit the side information available at DB1, i.e., $b_1$, in order to retrieve a new desired symbol $a_2$ from the second database (DB2) by mixing it with $b_1$. At this point the query to each database is symmetric, and there is no side information that remains unexploited. Thus the construction is complete. 
\begin{eqnarray*}
\begin{array}{|c|c|c|c|c|}\hline
\mbox{\tiny DB1}&\mbox{\tiny DB2}\\\hline
a_1&~~\\\hline
\end{array}\stackrel{(1)}{\longrightarrow}\begin{array}{|c|c|c|c|c|}\hline
\mbox{\tiny DB1}&\mbox{\tiny DB2}\\\hline
a_1,b_1& \\\hline
\end{array}\stackrel{(2)}{\longrightarrow}\begin{array}{|c|c|c|c|c|}\hline
\mbox{\tiny DB1}&\mbox{\tiny DB2}\\\hline
a_1,b_1&a_2 + b_1\\\hline
\end{array}
\end{eqnarray*}
Similarly, the queries for $\theta = 2$ are constructed as follows.
\begin{eqnarray*}
\begin{array}{|c|c|c|c|c|}\hline
\mbox{\tiny DB1}&\mbox{\tiny DB2}\\\hline
b_1&~~\\\hline
\end{array}\stackrel{(1)}{\longrightarrow}\begin{array}{|c|c|c|c|c|}\hline
\mbox{\tiny DB1}&\mbox{\tiny DB2}\\\hline
a_1,b_1& \\\hline
\end{array}\stackrel{(2)}{\longrightarrow}\begin{array}{|c|c|c|c|c|}\hline
\mbox{\tiny DB1}&\mbox{\tiny DB2}\\\hline
a_1,b_1&a_1 + b_2\\\hline
\end{array}
\end{eqnarray*}
Note that the application of Step (1) only introduces new terms that do not involve symbols from the desired message, whereas the application of Step (2) only introduces new terms that involve symbols from the desired message.

To see why this scheme is private, recall that $[a_1, a_2]$ are random permutations of two symbols from $W_1$ and $[b_1,b_2]$ are random permutations of two symbols from $W_2$. These permutations are known only to the user, and not to the databases. Therefore, regardless of whether $\theta=1$ or $\theta=2$, DB1 is asked for one randomly chosen symbol of each message, and DB2 is asked for a sum of a pair of randomly chosen symbols from each message. Since the permutations are uniform, all possible realizations are equally likely, and privacy is guaranteed.  A formal proof of privacy for the general setting appears in Section \ref{sec:proof}.

The scheme is correct, because each desired message symbol is either downloaded directly or as a sum with side information terms that are separately downloaded.

Finally, note that the download cost is $D = 3=\lceil \frac{L}{C}\rceil$, because $C = 2/3$ for this case. The rate achieved is $L/D=2/3=C$.

\subsubsection*{$K = 3$ Messages, $N = 2$ Databases, $L = N^{K-1}=4$ Symbols Per Message}
Let $[a_1, \cdots, a_4]$ represent a random permutation of $4$ $M$-ary symbols from message $W_1$. Similarly,  $[b_1, \cdots, b_4] $ and $[c_1, \cdots, c_4]$ are  random permutations of $4$ $M$-ary symbols each from messages $W_2,W_3$, respectively. The uniformly random and independent permutations are generated privately  by the user.
Suppose $\theta = 1$. The query generation algorithm proceeds as follows.
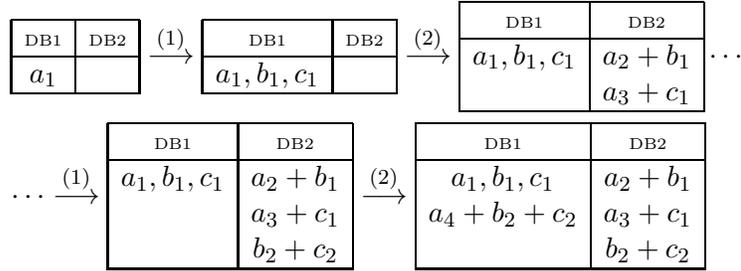
\begin{eqnarray*}
&&\begin{array}{|c|c|c|c|c|}\hline
\mbox{\tiny DB1}&\mbox{\tiny DB2}\\\hline
a_1&~~\\\hline
\end{array}\stackrel{(1)}{\longrightarrow}\begin{array}{|c|c|c|c|c|}\hline
\mbox{\tiny DB1}&\mbox{\tiny DB2}\\\hline
a_1,b_1,c_1&\\\hline
\end{array}\stackrel{(2)}{\longrightarrow}\begin{array}{|c|c|c|c|c|}\hline
\mbox{\tiny DB1}&\mbox{\tiny DB2}\\\hline
a_1,b_1,c_1&a_2 + b_1\\
&a_3+c_1\\\hline
\end{array}\cdots\\
&&\cdots\stackrel{(1)}{\longrightarrow}\begin{array}{|c|c|c|c|c|}\hline
\mbox{\tiny DB1}&\mbox{\tiny DB2}\\\hline
a_1,b_1,c_1&a_2 + b_1\\
&a_3+c_1\\
&b_2+c_2\\\hline
\end{array}\stackrel{(2)}{\longrightarrow}\begin{array}{|c|c|c|c|c|}\hline
\mbox{\tiny DB1}&\mbox{\tiny DB2}\\\hline
a_1,b_1,c_1&a_2 + b_1\\
a_4 + b_2 + c_2&a_3+c_1\\
&b_2+c_2\\
\hline
\end{array}\end{eqnarray*}
Again, note that the application of Step (1) only introduces new terms that do not involve symbols from the desired message, whereas the application of Step (2) only introduces new terms that involve symbols from the desired message.
The queries generated  when $\theta = 2, 3$ are as follows.
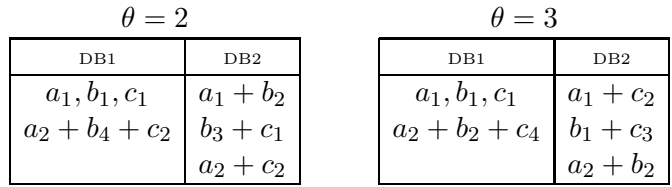
\begin{eqnarray*}
\begin{array}{cccc}
\theta=2&&&\theta=3\\
\begin{array}{|c|c|cc|c|c|}\hline
\mbox{\tiny DB1}& \mbox{\tiny DB2} \\\hline
a_1,b_1,c_1&a_1 + b_2\\
a_2 + b_4 + c_2&b_3+c_1\\
&a_2+c_2\\\hline
\end{array}
& & &
\begin{array}{|c|c|cc|c|c|}\hline
\mbox{\tiny DB1}& \mbox{\tiny DB2} \\\hline
a_1,b_1,c_1&a_1 + c_2\\
a_2 + b_2 + c_4&b_1+c_3\\
&a_2+b_2\\\hline
\end{array}
\end{array}
\end{eqnarray*}
Correctness is straightforward, privacy is ensured by message symmetry and random permutations, and the rate is $L/D = 4/7$ which matches the capacity $C$ for this case. The download achieved is $D=4$ symbols which is also optimal.

\subsubsection*{$K = 3$ Messages, $N = 3$ Databases, $L = N^{K-1}=9$ Symbols Per Message}
Let $[a_1, \cdots, a_9], [b_1, \cdots, b_9], [c_1, \cdots, c_9]$ be three  i.i.d. uniform permutations of symbols from messages $W_1, W_2, W_3$, respectively. 
The query generation algorithm for $\theta = 1$ proceeds as follows.

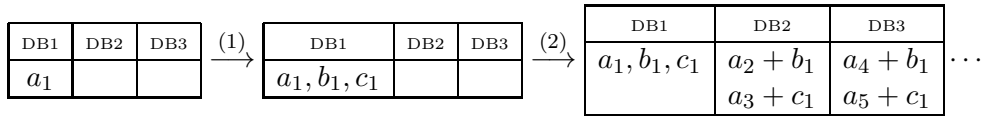
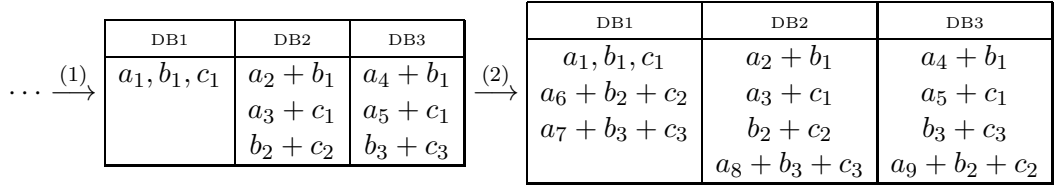
\begin{eqnarray*}
&&\begin{array}{|c|c|c|c|c|}\hline
\mbox{\tiny DB1}&\mbox{\tiny DB2}&\mbox{\tiny DB3}\\\hline
a_1&~~&~~\\\hline
\end{array}\stackrel{(1)}{\longrightarrow}\begin{array}{|c|c|c|c|c|}\hline
\mbox{\tiny DB1}&\mbox{\tiny DB2}&\mbox{\tiny DB3}\\\hline
a_1,b_1,c_1& & \\\hline
\end{array}\stackrel{(2)}{\longrightarrow}\begin{array}{|c|c|c|c|c|}\hline
\mbox{\tiny DB1}&\mbox{\tiny DB2}&\mbox{\tiny DB3}\\\hline
a_1,b_1,c_1&a_2 + b_1&a_4 + b_1\\
&a_3 + c_1&a_5 + c_1 \\\hline
\end{array}\cdots\\
&&\cdots\stackrel{(1)}{\longrightarrow}\begin{array}{|c|c|c|c|c|}\hline
\mbox{\tiny DB1}&\mbox{\tiny DB2}&\mbox{\tiny DB3}\\\hline
a_1,b_1,c_1&a_2 + b_1&a_4 + b_1\\
&a_3 + c_1&a_5 + c_1 \\
&b_2 + c_2&b_3 + c_3 \\\hline
\end{array}\stackrel{(2)}{\longrightarrow}\begin{array}{|c|c|c|c|c|}\hline
\mbox{\tiny DB1}&\mbox{\tiny DB2}&\mbox{\tiny DB3}\\\hline
a_1,b_1,c_1&a_2 + b_1&a_4 + b_1\\
a_6 + b_2 + c_2&a_3 + c_1&a_5 + c_1 \\
a_7 + b_3 + c_3&b_2 + c_2&b_3 + c_3 \\
& a_8 + b_3 + c_3&a_9+b_2 + c_2 \\
\hline
\end{array}
\end{eqnarray*}
Again, note that the application of Step (1) only introduces new terms that do not involve symbols from the desired message, whereas the application of Step (2) only introduces new terms that involve symbols from the desired message.
\noindent The scheme when $\theta = 2, 3$ is as follows.
\begin{eqnarray*}
\begin{array}{cccc}
\theta=2&&\theta=3\\
\begin{array}{|c|c|c|c|c|}\hline
\mbox{\tiny DB1}&\mbox{\tiny DB2}&\mbox{\tiny DB3}\\\hline
a_1,b_1,c_1&a_1 + b_2&a_1 + b_4\\
a_2 + b_6 + c_2&b_3 + c_1&b_5 + c_1 \\
a_3 + b_7 + c_3&a_2 + c_2&a_3 + c_3 \\
& a_3 + b_8 + c_3&a_2+b_9 + c_2 \\
\hline
\end{array}
&&
\begin{array}{|c|c|c|c|c|}\hline
\mbox{\tiny DB1}&\mbox{\tiny DB2}&\mbox{\tiny DB3}\\\hline
a_1,b_1,c_1&a_1 + c_2&a_1 + c_4\\
a_2 + b_2 + c_6&b_1 + c_3&b_1 + c_5 \\
a_3 + b_3 + c_7&a_2 + b_2&a_3 + b_3 \\
& a_3 + b_3 + c_8&a_2+b_2 + c_9 \\
\hline
\end{array}
\end{array}
\end{eqnarray*}

Correctness is straightforward, privacy is ensured by message symmetry and random permutations, and the rate is $L/D = 9/13$ which matches the capacity $C$ for this case. The download achieved is $D=13$ symbols which is also optimal.


Next we go beyond the simple examples to the general $N,K$ setting. Let us start by introducing some new definitions and notation, some of which is needed only to suppress those aspects of the general setting that are notationally cumbersome but otherwise inconsequential.

\subsection{Definitions and Additional Notation}\label{sec:def}
\noindent{\bf [$U_k$]} For all $k\in[1:K]$, define\footnote{The $U_k$ symbols will eventually be mapped to random permutations of message $W_k$ symbols. We use $[U_k(l)]$ instead of, say $[a_l], [b_l]$ as in the examples, because while the latter notation is more clear, it does not generalize to $K$ messages.}
ordered tuples
 \begin{eqnarray}
 U_k\triangleq [ U_k(1), U_k(2),\cdots, U_k(N^{K-1}) ]
 \end{eqnarray}
 
 \bigskip
 
\noindent{\bf [$k$-sums, Types]} We  use the terminology {\bf $k$-sum} to denote an expression representing the  sum of $k$ distinct  variables, each drawn from a \emph{different} $U_i$ tuple, i.e., $U_{i_1}(j_1)+U_{i_2}(j_2)+\cdots+U_{i_k}(j_k)$, where $i_1,i_2,\cdots, i_k\in[1:K]$ are all \emph{distinct} indices. Furthermore, we will define such a $k$-sum to be of {\bf type} $\{i_1,i_2,\cdots,i_k\}$, or $i_{[1:k]}$ in our compact notation. If $q$ represents a $k$-sum, the function $\m{type}(q)$ returns its type. Denote $\mathcal{T}_k$ as the set of all possible types of a $k$-sum, i.e., all possible choices of $k$ distinct indices in $[1:K]$. Note that $|\mathcal{T}_k| = \binom{K}{k}$.

\bigskip

The next two  items are introduced to facilitate a compact notation. The first of these is a function, $\new(\cdot)$, which will allow us to suppress unimportant details about symbol indices.

\bigskip

\noindent{\bf [The $\new(\cdot)$ Function]}
 For any ordered tuple $U$, let $\new(U)$ be a function that, starting with $U(1)$,  returns the ``next" element in $U$ each time\footnote{We will deal with $N^{K-1}$-tuples and the algorithms will guarantee that the $\new(.)$ function is not called more than $N^{K-1}$ times for the same tuple.} it is called with the same tuple $U$ as its argument. So, for example, the following sequence of  calls to this function:  $\new(U_2), \new(U_1),\new(U_1), \new(U_1)+ \new(U_2)$ will produce 
$U_2(1), U_1(1), U_1(2), U_1(3)+ U_2(2)$
as the output. 

\bigskip 

\noindent{\bf [Ordered Access to Elements of a Set]}
 In a similar spirit, for any set $A$, when accessing its elements (e.g., in an algorithm), we will use the notation $\overrightarrow{A}$  to indicate that the elements of $A$ are to be accessed in some specified order, the details of which are not significant, because all  ordering rules will produce  (possibly different) optimal PIR schemes. Let us assume by default that the ordering is the natural lexicographic  increasing order. For example,  $\overrightarrow{[1:K]}$ refers to increasing order of integers $1$ through $K$. $\overrightarrow{\mathcal{T}_k}$ denotes that the types, i.e., the $\{i_1,i_2,\cdots, i_k\}$ terms in $\mathcal{T}_k$ are individually sorted and then accessed in lexicographic increasing order. For a set $Q$ that is comprised of various $k$-sums the notation $\overrightarrow{Q}$ represents that the order in which the elements are accessed is, first in increasing order of $k$, then within the same $k$ in increasing order of type, and then for multiple instances of the same type the elements are accessed in increasing order of the $j$ index of the $U_i(j)$ with the smallest $i$. Some examples of this notation:
 \begin{align}
 \bigcup_{k\in\overrightarrow{[1:2]}}\{U_1(k)+\new(U_2)\}&=\{U_1(1)+U_2(1), U_1(2)+U_2(2)\}\\
 \bigcup_{q\in\overrightarrow{Q}}\{q+\new(U_1)\}&=\{U_1(1)+U_2(4),U_1(2)+U_2(2)+U_3(3),U_1(3)+U_2(3)+U_3(2)\}
 \end{align}
where $Q=\{U_2(2)+U_3(3), U_2(4), U_2(3)+U_3(2)\}$, so that $\overrightarrow{Q}$ denotes that the terms of $Q$ are accessed in the order $U_2(4), U_2(2)+U_3(3), U_2(3)+U_3(2)$. 

\bigskip

\noindent{\bf [The $\mbox{Count}(\cdot)$ and $\mbox{Max}(\cdot)$ Functions]}
$\mbox{Count}(Q,i_{[1:k]})$ denotes the number of $k$-sums  of type $\{i_1,i_2,\cdots,i_k\}$ that are present in $Q$
\begin{eqnarray}
\mbox{Count}(Q, i_{[1:k]}) \define |\{q :~ q \in Q, \m{type}(q) =i_{[1:k]}\}|,
\end{eqnarray}
$\mbox{Max}(Q, k)$ denotes the maximum of the number of $k$-sums of the same type in $Q$, with the maximization being across all types of $k$-sums,
\begin{eqnarray}
\mbox{Max}(Q, k) \define \max_{i_{[1:k]} \in \mathcal{T}_k}  \mbox{Count}(Q, i_{[1:k]})
\end{eqnarray}

\noindent{\bf [Message Symmetry]}
Message symmetry is defined as the condition that  $\forall k\in[1:K]$, $Q$ contains equal number of $k$-sums for every type $\{i_1,i_2,\cdots,i_k\}\in \mathcal{T}_k$.
\begin{align}
&& \mbox{Count}(Q, i_{[1:k]}) = \mbox{Count}(Q, i'_{[1:k]}), ~~~\forall i_{[1:k]},i'_{[1:k]}\in\mathcal{T}_k \label{eq:msym}
\end{align}

\subsection{Two Subroutines}\label{sec:twosub}
For the sake of clarity, here we separately present the two procedures needed to implement the message symmetry and side-information exploitation steps, which will ultimately be incorporated into the overall query generation algorithm.
\subsubsection{Algorithm $(1)$: Achieving Message Symmetry (M-Sym Algorithm)}
The algorithm takes as input a set $Q$ comprised of various $k$-sums, and produces as output a set $Q^*$ comprised of additional terms that need to be included in $Q$ to make it message symmetric, i.e., $Q\cup Q^*$ satisfies message symmetry. For each $k\in[1:K]$, and for each type $i_{[1:k]}\in\mathcal{T}_k$, the algorithm checks if there are $\m{Max}(Q,k)$ instances of that type, and if not, then it generates as many new instances as necessary to bring up the number of instances of that type  to $\m{Max}(Q,k)$.
\allowdisplaybreaks
\begin{algorithm}[H]
\caption{M-Sym Algorithm.}
\label{msym}
\begin{algorithmic}[1]{}
\STATE{{\bf Input:} $Q$}
\STATE{{\bf Output:} $Q^*$}
\STATE{Initialize: $Q^* \leftarrow \emptyset$.} 
\FOR
{${k}=1:K$} 
\FOR
{\textbf{each} $i_{[1:k]}\in \overrightarrow{\mathcal{T}_k} $}
\IF{$\mbox{Count}(Q, i_{[1:k]}) < \mbox{Max}(Q, k) $}
\FOR
{$i = 1 : \mbox{Max}(Q, k) - \mbox{Count}(Q, i_{[1:k]})$}
\STATE{$$Q^* \leftarrow Q^* \cup \{\new(U_{i_1}) + \new(U_{i_2}) + \cdots + \new(U_{i_k})  \}$$} 
\ENDFOR \m{ ($i$)}
\ENDIF
\ENDFOR \m{ ($i_{[1:k]}$)}
\ENDFOR { ($k$)}
\end{algorithmic}
\end{algorithm}

Note that $Q\cup Q^*$ satisfies message symmetry because for all types $i_{[1:k]} \in \mathcal{T}_k$, $\mbox{Count}(Q\cup Q^*, i_{[1:k]}) = \mbox{Max}(Q\cup Q^*,k)=\mbox{Max}(Q,k)$.


\subsubsection{Algorithm $(2)$: Exploiting Side Information (Exploit-SI Algorithm)}
Algorithm (2) formalizes the side information exploitation step. This algorithm takes as input $N$ query sets $Q_1, Q_2, \cdots, Q_N$, which are comprised of side-information terms, i.e., terms that do not contain any desired message symbols, i.e., $\forall n\in[1:N]$ and $\forall q\in Q_n$, $\theta\notin \m{type}(q)$ and which have not previously been exploited.  The algorithm produces $N$ sets $Q'_1, Q'_2, \cdots, Q'_N$ as output. $Q'_n, n \in [1:N]$ is built by combining each element $q$ in $Q'_1, \cdots, Q'_{n-1}, Q'_{n+1}, \cdots, Q'_{N}$ with a ``$\new$" variable $U_\theta$ (which corresponds to a desired message symbol). 
\allowdisplaybreaks
\begin{algorithm}[H]{}
\caption{Exploit-SI Algorithm.}
\label{exploit}
\begin{algorithmic}[1]{}
\STATE{{\bf Input:} $Q_1,Q_2,\cdots, Q_N$}
\STATE{{\bf Output:} $Q'_1,Q'_2,\cdots, Q'_N$}
\STATE{Initialize: All output  are initialized as null sets.} 
\FOR
{${n}=1:N$} 
\FOR
{
$n' = 1:N$ \textbf{and} $n' \neq n$}
\FOR
{\textbf{each} $q \in \overrightarrow{Q_{n'}}$
}
\STATE{$$Q'_n \leftarrow Q'_{n} \cup \{\new(U_{\theta}) + q  \}$$} \ENDFOR \m{ ($q$)} 
\ENDFOR \m{ (${n}'$)}
\ENDFOR { (${n}$)}
\end{algorithmic}
\end{algorithm}



\bigskip

\subsection{A Deterministic Query Generation Algorithm (Q-Gen Algorithm)}\label{sec:Q-Gen}
We now proceed to a  query generation algorithm.\footnote{Note that this is not the final step in the query generation. The output of this deterministic algorithm is in terms of the  $U_k$ variables. The final step, to be presented in Section \ref{sec:qfinal}, maps $U_k$ variables to private random permutations of $W_k$ variables, to produce the random queries that are then sent to the databases.} The algorithm  produces $N$ query sets $Q(\m{DB},\theta)$, for all $\m{DB}\in[1:N]$ as functions of $\theta$. For internal book-keeping in the algorithm, we will partition each query set into $K$ subsets called blocks, such that block $k\in[1:K]$ contains only $k$-sums. Further we will partition each block into two subsets denoted by $\mathcal{I}$ and $\mathcal{M}$ such that the $\mathcal{M}$ partition contains only those types of $k$-sums which involve variables  from $U_\theta$, and the $\mathcal{I}$ partition contains the remaining $k$-sums which do not involve the $U_\theta$  variables.

As  in the simple  examples presented earlier, for all $\m{DB} \in [1:N], \theta \in [1:K]$, the query sets $Q(\m{DB}, \theta)$  are built starting only from a single element in $Q(1, \theta)$, which is the first desired message symbol $U_\theta$, and then evolves through iterative application of the M-Sym and Exploit-SI sub-routines. Note that the memory of calls to the  $\new(\cdot)$ function is assumed to be global, i.e., the memory is not reset when the sub-routines are called. Similarly, $\theta$ is assumed to be available to the sub-routines as a global variable.

\allowdisplaybreaks
\begin{algorithm}[H]
\caption{Q-Gen Algorithm.}
\label{alg1}
\begin{algorithmic}[1]{}
\STATE{{\bf Input:} $\theta$}
\STATE{{\bf Output:} $Q({1},\theta), \cdots, Q(N,\theta)$}
\STATE{Initialize: All query sets are initialized as null sets. Also initialize $\m{Block}\leftarrow 1$;} 
\STATE{\begin{align}Q(\m{1},\theta,\m{Block},\mathcal{M}) &\leftarrow\{\new(U_\theta)\}\\~
Q(\m{1},\theta,\m{Block},\mathcal{I}) &\leftarrow \m{\textbf{\color{blue} M-Sym}}(Q(\m{1},\theta,\m{Block},\mathcal{M})) 
 \\
\forall \m{DB} \in [2:N], ~~~Q(\m{DB},\theta,\m{Block},\mathcal{M})\leftarrow \emptyset, ~&Q(\m{DB},\theta,\m{Block},\mathcal{I}) \leftarrow \emptyset, 
\end{align}
}

\FOR
{$\m{Block}=2:K$} 
\STATE{
$$( Q({1},\theta,\m{Block},\mathcal{M}), \cdots,  Q({N},\theta,\m{Block},\mathcal{M})) \leftarrow \m{\textbf{\color{blue} Exploit-SI}}( Q({1},\theta,\m{Block}-1,\mathcal{I}), \cdots,  Q({N},\theta,\m{Block}-1,\mathcal{I}))$$
}

\FOR
{$\m{DB}=1:N$}
\STATE{$$ Q(\m{DB},\theta,\m{Block},\mathcal{I}) \leftarrow \m{\textbf{\color{blue} M-Sym}}(Q(\m{DB},\theta,\m{Block},\mathcal{M}))
$$
}
\ENDFOR \m{ (DB)}
\ENDFOR \m{ (Block)}
\FOR{$\m{DB}=1:N$}
\STATE {
$
Q(\m{DB},\theta)\leftarrow\bigcup_{\m{\tiny Block}\in[K]} \big(Q(\m{DB},\theta,\m{Block},\mathcal{I})\cup Q(\m{DB},\theta,\m{Block},\mathcal{M})\big) \label{ni}
$}
\ENDFOR \m{ (DB)}
\end{algorithmic}
\end{algorithm}
Based on Algorithm \ref{alg1}, we have two immediate observations.
\begin{enumerate}
\item Consider the number of instances with type $\{i_1, \cdots, i_{k-1},\theta\}$ in $Q(\m{DB},\theta, k,\mathcal{M})$, i.e., $$\m{Count}(Q(\m{DB},\theta, k,\mathcal{M}), \{ i_1, \cdots, i_{k-1},\theta\}).$$ 
$Q(\m{DB},\theta, k,\mathcal{M})$ is produced in {\color{black}Step 6 of Algorithm (3)} as one of the outputs of the $\m{Expoit-SI}$ algorithm. From  {\color{black}Step 7} of the $\m{Expoit-SI}$ algorithm, we know that the instances with type $\{i_1, \cdots, i_{k-1},\theta \}$ in $Q(\m{DB},\theta, k,\mathcal{M})$ are produced by combining a new variable from $U_{\theta}$ with each element of type $\{i_1, \cdots, i_{k-1}\}$ in $Q(\m{DB}',\theta, k-1,\mathcal{I}), \m{DB}' \neq \m{DB}$, i.e.,
\begin{align}
& \m{Count}(Q(\m{DB},\theta, k,\mathcal{M}), \{i_1, \cdots, i_{k-1},\theta\}) = \sum_{\m{\tiny DB}' \neq \m{\tiny DB}} \m{Count}(Q(\m{DB}',\theta, k-1,\mathcal{I}), \{i_1, \cdots, i_{k-1}\}) \notag\\
&~~~ \forall  \m{DB} \in [1:N], \theta \in [1:K], k \in [2:K], \{i_1, \cdots, i_{k-1}\} \in \mathcal{T}_{k-1} \label{eq:des}
\end{align}

\item From {\color{black}Step 4} and {\color{black}Step 8} of Algorithm (3), we know that $Q(\m{DB},\theta,k,\mathcal{I}) \cup Q(\m{DB},\theta,k,\mathcal{M}), \forall k \in [1:K]$ satisfies message symmetry (\ref{eq:msym}).
\end{enumerate}

\subsubsection*{Structure of $Q(\m{DB}, \theta)$}
Key properties of $Q(\m{DB}, \theta)$ are summarized in the following lemma.
\begin{lemma}\label{lemma:property}
$Q(\m{DB}, \theta)$ produced by Algorithm \ref{alg1} satisfies the following properties.
\begin{enumerate}
\item $Q(\m{DB}, \theta), \forall \m{DB} \in [1:N], \theta \in [1:K]$ is a union of $K$ disjoint sets (called ``blocks''), that are indexed by $k \in [1:K]$. Block $k$ only contains $k$-sums. For any type $i_{[1:k]}\in \mathcal{T}_k$, block $k$ of $Q(\m{DB}, \theta)$ contains $v(\m{DB}, k)$ instances of type $i_{[1:k]}$, where $v(\m{DB}, k)$ is a function only of $\m{DB}, k$.


\item $\forall i\in[1:K]$, if $U_i(j)$ and $U_i(j')$  appear anywhere in the same $Q(\m{DB},\theta)$ then $j\neq j'$.

\item Exactly $v(\m{DB}) \define \sum_{k=1}^K v(\m{DB},k) \binom{K-1}{k-1}$ distinct variables for each $U_i, i \in [1:K]$ appear in $Q(\m{DB},\theta)$.
\end{enumerate}
\end{lemma}

{\it Proof:}
\begin{enumerate}
\item Block $k, k \in [1:K]$ of $Q(\m{DB}, \theta)$ is the set $Q(\m{DB},\theta, k,\mathcal{I})\cup Q(\m{DB},\theta, k,\mathcal{M})$, which satisfies message symmetry based on Observation 2. From {\color{black} Step 4 of Algorithm (3)}, we know that Block 1 only contains 1-sums. From {\color{black}Steps 6 and 8}, we know that the type of each instance in Block $k, k \in [2:K]$ contains one more variable than that of any instance in Block $k-1$. Therefore, by induction, Block $k$ only contains $k$-sums. As each Block $k$ satisfies message symmetry, we have
\begin{align}
\m{Count}(Q(\m{DB},\theta, k,\mathcal{M}), \{ i_1, \cdots, i_{k-1},\theta\}) &=  \m{Max}(Q(\m{DB},\theta), k)  \\
\m{Count}(Q(\m{DB},\theta, k-1,\mathcal{I}), \{i_1, \cdots, i_{k-1}\})  &= \m{Max}(Q(\m{DB},\theta), k-1) 
\end{align}
and (\ref{eq:des}) reduces to
\begin{align}
\m{Max}(Q(\m{DB},\theta), k)  &= \sum_{\m{\tiny DB}' \neq \m{\tiny DB}} \m{Max}(Q(\m{DB}',\theta), k-1)  
\label{vv}
\end{align}
Combined with the fact that $\m{Max}(Q(1,\theta), 1) = 1, \m{Max}(Q(\m{DB},\theta), 1)  = 0, \forall \m{DB} \in [2:N]$ (obtained from {Step 4 of Algorithm (3)}), we conclude that $\m{Max}(Q(\m{DB},\theta), k)$ depends only on $\m{DB}$ and $k$. 
Therefore, $v(\m{DB}, k) = \m{Max}(Q(\m{DB},\theta), k)$ and $v(\m{DB}, k)$ is a function of only $\m{DB}, k$.

\item Fix any database $\m{DB}$. Consider the case where $i = \theta$ first. Note that desired variables only appear in $Q(\m{DB}, \theta, \m{Block}, \mathcal{M})$. From Step 4 and Step 6 in Algorithm \ref{alg1}, we see that the desired variables, i.e., the $U_\theta$ variables appear only through the $\new(U_\theta)$ function so that each of them has a distinct index. Next, consider the non-desired variables, $U_{i}, i \neq \theta$, which either appear in Steps 4 and 8 through the $\new(U_k)$ function or appear in Step 6 which in turn come from $Q(\m{DB}, \theta, \m{Block}-1, \mathcal{M})$ and each of them was introduced once through the $\new(U_k)$ function and used exactly once. Therefore, these $U_k$ variables also have distinct indices within $Q(\m{DB}, \theta)$. 

\item This claim follows directly from the previous two claims.
Note that we have shown that all variables from $U_i$ are distinct, so $v(\m{DB})$ is equal to the number of times that variables in $U_i$ appear in $Q(\m{DB}, \theta)$. In the $k$-th block, $Q(\m{DB}, \theta)$ contains $v(\m{DB},k)$ instances of $k$-sums of each type and there are $\binom{K}{k-1}$ types of $k$-sums that include $i$. Therefore, the number of instances of tuple $U_i$ in block $k$ is $v(\m{DB},k) \binom{K-1}{k-1}$. Summing over all $K$ blocks, we have $v(\m{DB}) = \sum_{k=1}^K v(\m{DB},k) \binom{K-1}{k-1}$. 
\end{enumerate}
\hfill\QED

According to Lemma \ref{lemma:property} the query sets $Q(\m{DB},\theta)$ are comprised of $K$ blocks,  the $k$-th block contains $v(\m{DB},\theta)$ instances of every possible type of $k$-sum, and no $U_i(j)$ variable appears more than once in $Q(\m{DB},\theta)$. Therefore, the structure of the query set may be summarized in the following corollary.

\begin{corollary}  \label{cor:structure}
Given $\m{DB},\theta$, for every $U_k$, $k\in[1:K]$, there exists its permutation $\underline{U}_k$ that depends only on $\m{DB},\theta,k$,
\begin{eqnarray}
\underline{U}_{k}&\triangleq& \lambda_{\m{\tiny DB},\theta, k}(U_k)
\end{eqnarray}
such that $Q(\m{DB}, \theta)$ can be expressed as
\begin{eqnarray}
Q(\m{DB}, \theta) = \bigcup_{k\in\overrightarrow{[1:K]}} ~\bigcup_{i_{\tiny [1:k]} \in \overrightarrow{\mathcal{T}_k}} ~~~\bigcup_{l=1}^{v(\m{\tiny DB},k)} \{\new(\underline{U}_{i_1})+\new(\underline{U}_{i_2})+\cdots+\new(\underline{U}_{i_k})\}
\end{eqnarray}
\end{corollary}


{\it Remark:} As an example, consider the  example with $K = 3, N = 3, L = 9$ that was presented earlier in Section \ref{sec:ex}. Suppose $\m{DB} = 2, \theta = 3$. The query $Q(\m{DB}, \theta) = Q(2,3)$ is reproduced as follows.
\begin{eqnarray*}
Q(2, 3) = \{a_2 + b_2, a_1 + c_2, b_1 + c_3, a_3 + b_3 + c_8\},
\end{eqnarray*}
which 
can be equivalently written in the form in Corollary \ref{cor:structure} by setting 
\begin{eqnarray}
\lambda_{2,3, 1}(U_1) &=& (a_2, a_1, a_3, a_4, a_5, a_6, a_7, a_8, a_9)\\
\lambda_{2,3, 2}(U_2) &=& (b_2, b_1, b_3, b_4, b_5, b_6, b_7, b_8, b_9) \\
\lambda_{2,3, 3}(U_3) &=& (c_2, c_3, c_8, c_1, c_4, c_5, c_6, c_7, c_9)
\end{eqnarray}
Note that here $U_1 = [a_1, \cdots, a_9], U_2 = [b_1, \cdots, b_9], U_3 = [c_1, \cdots, c_9]$.

\subsection{
Mapping to Message Symbols  to Produce $Q_{\m{\tiny DB}}^{[\theta]}$}\label{sec:qfinal}
To produce the actual query sent to the databases, we map the $U_k(i)$ variables to message symbols. This mapping is specified by $K$ privately chosen permutations $\gamma_1, \gamma_2, \cdots, \gamma_K$, each of which is uniformly random over all possible $(N^{K-1})!$ permutations over the index set $[1:N^{K-1}]$ and the permutations are independent of each other and of $\theta$. Specifically, $U_k(i)$ is replaced with $W_k(\gamma_k(i))$, $\forall k\in[1:K], i\in[1:N^{K-1}]$. This operator is denoted by $\Gamma$ . For example, $\Gamma(\{U_1(2),U_3(4)+U_5(6)\})=\{W_1(\gamma_1(2)),W_3(\gamma_3(4))+W_5(\gamma_5(6))\}$. After this random mapping is applied to ${Q}(\m{DB},\theta)$, we obtain the actual query set  $Q_{\m{\tiny DB}}^{[\theta]}$ that is  sent to database $\m{DB}$.
\begin{eqnarray}
Q_{\m{\tiny DB}}^{[\theta]}&=& ``\Gamma\big({Q}(\m{DB},\theta)\big)"
\end{eqnarray}
We use the double-quotes notation around a symbol to represent the \emph{query} about its realization.  For example, while $W_1(1)$ is the realization of one message symbol, in our notation ``$W_1(1)$" only represents the \emph{question}: ``what is the value of $W_1(1)$?" $Q_{\m{\tiny DB}}^{[\theta]}$ is a (unordered) set and the questions in the set are sent in an order that is independent of $\theta$ (say, uniformly random) to the databases.

\subsection{Proof of Correctness, Privacy and Optimality}\label{sec:proof}
We prove that the achievable scheme is correct, private and optimal in the following two lemmas.

\begin{lemma}
The PIR scheme constructed through the {\bf Q-Gen} Algorithm is correct, i.e., it satisfies (\ref{corr}). The message size is $L = N^{K-1}$ and the download cost is optimal, $D = \frac{L}{C}$. 
\end{lemma}
{\it Remark:} $\frac{L}{C}$ is an integer, so that $D_L = \lceil \frac{L}{C} \rceil = \frac{L}{C}$.\\

{\it Proof:} 
Note that all desired message symbols  are either retrieved directly with no interference or  they appear with interference $q$ that is downloaded separately from another database so it can be subtracted to retrieve the desired symbols. Therefore, all the desired message symbols are retrievable and the correctness constraint (\ref{corr}) is satisfied.

In order to compute the message size and download cost, we proceed as follows. Using (\ref{vv}), we have
\begin{align}
v(1, 1) &= 1\\
 v(\m{DB}, 1) &= 0, \forall \m{DB} \in [2:N] \label{in} \\
v(\m{DB}, k) &= \sum_{\m{\tiny DB}' \neq \m{\tiny DB}} v(\m{DB}', k-1), \forall k \in [2:K] \label{use}\\
\Rightarrow  v(2, k) = \cdots &= v(N, k), \forall k \in [2:K] \label{sym}
\end{align}
For all $k\in[2:K]$,
\begin{align}
v(1,k) &\stackrel{(\ref{use})(\ref{sym})}{=} (N-1) v(2,k-1) \label{v1}\\
v(2,k) &\stackrel{(\ref{use})(\ref{sym})}{=}  v(1,k-1) +(N-2) v(2,k-1) \label{v2}\\
\Rightarrow  v(1,k) + (N-1) v(2,k) &\stackrel{(\ref{v1})(\ref{v2})}{=}  (N-1) v(2,k-1) + (N-1) \Big( v(1,k-1) +(N-2) v(2,k-1)  \Big) \notag\\
&= (N-1) \Big( v(1,k-1) + (N-1) v(2,k-1) \Big) \\
&= \cdots \\
&= (N-1)^{k-1} \Big( v(1,1) + (N-1) v(2,1)  \Big) \\
&\stackrel{(\ref{in})}{=} (N-1)^{k-1} \label{fg}
\end{align}

From Lemma \ref{lemma:property}, we have shown that there are $v(\m{DB}) = \sum_{k=1}^K v(\m{DB},k) \binom{K-1}{k-1}$ desired variables in each $Q(\m{DB}, \theta)$. Note that desired variables all appear through $\new(U_\theta)$ so that they are distinct across databases. Thus the message size (the total number of desired symbols that are retrieved) is
\begin{align}
L &= \sum_{\m{\tiny DB} = 1}^N \sum_{k=1}^K v(\m{DB},k) \binom{K-1}{k-1} \stackrel{(\ref{sym})}{=}  \sum_{k=1}^K \Big(v({1},k)+ (N-1) v(2,k) \Big)\binom{K-1}{k-1} \\
&\stackrel{(\ref{fg})}{=} \sum_{k=1}^{K} (N-1)^{k-1} \binom{K-1}{k-1} = \sum_{k=0}^{K-1} (N-1)^k \binom{K-1}{k} = (N-1 +1)^{K-1} = N^{K-1} \label{le}
\end{align} 
We next compute the download cost and show that the achieved download cost is optimal, i.e., $D = \frac{L}{C}$.
The $k$-th block of $Q(\m{DB},\theta)$ contains $v(\m{DB},k)$ instances of $k$-sums of each possible type, and there are $\binom{K}{k}$ possible types of $k$-sums. Therfore, the cardinality of $Q(\m{DB},\theta)$ is $\sum_{k=1}^K v(\m{DB},k) \binom{K}{k}$. Summing over all databases, we have
\begin{align}
D &= \sum_{\m{\tiny DB} = 1}^N \sum_{k=1}^K v(\m{DB},k) \binom{K}{k}\\
& \stackrel{(\ref{sym})}{=}  \sum_{k=1}^K \Big( v(1,k) + (N-1) v(2,k) \Big)\binom{K}{k} \\&\stackrel{(\ref{fg})}{=} \sum_{k=1}^{K} (N-1)^{k-1} \binom{K}{k} \\
&= \sum_{k=1}^{K-1} (N-1)^{k-1} \binom{K}{k} +(N-1)^{K-1}\\
&= \sum_{k=1}^{K-1}  (N-1)^{k-1} \left[ \binom{K-1}{k-1} + \binom{K-1}{k}\right]+(N-1)^{K-1}\\
&= \sum_{k=1}^{K}  (N-1)^{k-1}\binom{K-1}{k-1} + \sum_{k=1}^{K-1}  (N-1)^{k-1} \binom{K-1}{k}  \\
& \stackrel{(\ref{le})}{=} N^{K-1} + \sum_{k=1}^{K-1}  (N-1)^{k-1} \binom{K-1}{k}  \\
&= L + \frac{1}{N-1} \sum_{k=1}^{K-1}  (N-1)^{k} \binom{K-1}{k} = L + \frac{1}{N-1} \left[ \sum_{k=0}^{K-1}  (N-1)^{k} \binom{K-1}{k} - 1\right] \\
&= L + \frac{1}{N-1}(N^{K-1} - 1) = L + N^{K-1}\left( \frac{\frac{1}{N} - \frac{1}{N^K} }{1 - \frac{1}{N}}\right) = L \left(\frac{1 - \frac{1}{N^K}}{1- \frac{1}{N}}\right) = \frac{L}{C}
\end{align}
\hfill\QED

\begin{lemma}
The PIR scheme constructed through the {\bf Q-Gen} Algorithm is private, i.e., it satisfies (\ref{privacy}).
\end{lemma}
{\it Proof:}
From Corollary \ref{cor:structure}, we know that $Q(\m{DB},\theta)$  depends on $\theta$ only through the permutation functions $\lambda_{\m{\tiny DB},\theta,k}(U_k)$, for each $k\in[1:K]$.
But, $U_k$ are uniform permutations of message symbols, $U_k= \gamma_k(W_k)$. Because any permutation of a uniform permutation is also uniform,
\begin{eqnarray}
\lambda_{\m{\tiny DB},\theta,k}(\gamma_k(W_k)))&\sim&\gamma_k(W_k).
\end{eqnarray}
Furthermore, because $\gamma_1, \gamma_2, \cdots, \gamma_j$ are independent,
\begin{align}
(\lambda_{\m{\tiny DB},\theta,1}(\gamma_1(W_1))),\lambda_{\m{\tiny DB},\theta,2}(\gamma_2(W_2))),\cdots,\lambda_{\m{\tiny DB},\theta,K}(\gamma_K(W_K))))&\sim&(\gamma_1(W_1), \gamma_2(W_2),\cdots, \gamma_K(W_K))
\end{align}
Since $Q(\m{DB},\theta)$ is a function of $(\lambda_{\m{\tiny DB},\theta,1}(\gamma_1(W_1))),\lambda_{\m{\tiny DB},\theta,2}(\gamma_2(W_2))),\cdots,\lambda_{\m{\tiny DB},\theta,K}(\gamma_K(W_K))))$, which is identically distributed for all $\theta\in[1:K]$, $Q(\m{DB},\theta)$ is also identically distributed for all $\theta\in[1:K]$. Thus condition (\ref{privacy}) is satisfied and the scheme is private.

\hfill\QED

\section{Proof of Theorem \ref{thm:download}: Achievability for Arbitrary $L$}\label{proof:achLgen}

The optimal PIR scheme is a combination (analogous to time sharing arguments in channel capacity studies) of the capacity achieving scheme with message size $L = N^{K-1}$ that was presented in the previous section, and a PIR scheme from \cite{Shah_Rashmi_Kannan} (see the remark on replicated storage above Section V of \cite{Shah_Rashmi_Kannan})  which is related to blind interference alignment as noted in \cite{Sun_Jafar_BIAPIR} (see the discussion section of \cite{Sun_Jafar_BIAPIR}). Since the main objective of \cite{Shah_Rashmi_Kannan} is  PIR  with distributed storage, the scheme that we need is recovered as an implicit special case of  \cite{Shah_Rashmi_Kannan} (when replication coding is used across the databases). To make the scheme explicit, we restate this result in the following theorem. 

\begin{theorem}\label{lemma:bia}\cite{Shah_Rashmi_Kannan}
For PIR with $N \geq 2$ databases, each storing $K \in \mathbb{N}$ messages, each message comprised of $L= N-1$ symbols from $M$-ary alphabet, $M \in \mathbb{N}/\{1\}$, where the downloads are comprised of symbols from the same $M$-ary alphabet, the download cost $D = N = L+1$ $M$-ary symbols is achievable. 
\end{theorem}
While the scheme is implicitly contained in \cite{Shah_Rashmi_Kannan}, for the sake of completeness we give an explicit proof of Theorem \ref{lemma:bia} in Section \ref{sec:biaproof}. We also note that the binary alphabet ($M = 2$) case is considered recently  in \cite{Blackburn_Etzion_Paterson} (see Construction 1 of \cite{Blackburn_Etzion_Paterson}).

\subsection{Examples}
To convey the main ideas let us start with some examples for small values of $K, N, L$. The idea of constructing the optimal achievable scheme is to greedily use the most efficient PIR scheme (the capacity achieving scheme) repeatedly, and when the number of remaining symbols per message is less than required, we turn to the next most efficient scheme (the scheme in Theorem \ref{lemma:bia}), and continue to use the scheme in Theorem \ref{lemma:bia} with possibly smaller and smaller message sizes until all symbols are considered.

\subsubsection*{$K = 2$ Messages, $N = 2$ Databases, $L = 3$ Symbols Per Message} 
We show that the download cost $D = \lceil \frac{L}{C} \rceil = \lceil 3/(2/3) \rceil = 5$ symbols is achievable. The scheme is as follows. For each message, divide the $L = 3$ message symbols into two groups, where the first group is comprised of 2 symbols and the second group is comprised of 1 symbol. For the first group, we use the capacity achieving scheme with message length $N^{K-1} = 2$ so that the download cost achieved is $2/C = 3$ symbols. For the second group, we use the scheme described in Theorem \ref{lemma:bia} so that the download cost achieved is $N = 2$ symbols. Adding the two, the overall download cost is $D = 5$ symbols, as desired. 

\subsubsection*{$K = 3$ Messages, $N = 3$ Databases, $L = 16$ Symbols Per Message} 
We show that the download cost $D = \lceil \frac{L}{C} \rceil = \lceil 16/(9/13) \rceil = 24$ symbols is achievable. The scheme is as follows. For each message, divide the $L = 16$  symbols into three groups, where the first group is comprised of $9$ symbols, the second group is comprised of $6$ symbols and the third group is comprised of 1 symbol. For the first group, we use the capacity achieving scheme with message length $N^{K-1} = 9$ so that the download cost achieved is $9/C = 13$ symbols. Note that the second group only has $6$ symbols per message so that we can not use the capacity achieving scheme and we turn to the scheme in Theorem \ref{lemma:bia}.  For the second group, we further divide the 6 symbols to 3 sub-groups, each of which is comprised of 2 symbols. For each sub-group, we use the scheme described in Theorem \ref{lemma:bia} with $N = 3$ databases, so that the download cost per sub-group is $N = 3$ symbols. In total, the download cost for the second group is $9$ symbols. Note now that the third group only has $1$ symbol per message so that even the scheme for the second group does not apply and we turn to the same class of scheme but with shorter (matching) message length. For the third group, we use the scheme described in Theorem \ref{lemma:bia} with $N' = 2$ databases (say, the first two databases) and message size $L' = 1$ symbol (matching the size of the third group), so that the download cost achieved is $N' = 2$ symbols. Adding the download cost of the three groups up, the overall download cost is $D = 13 + 9 + 2 = 24$ symbols, as desired.

\subsection{Description of Achievable Scheme for Arbitrary $L$}
We now describe the general achievable scheme for arbitrary $L$, following the examples presented above. We first fully use the capacity achieving scheme with message size $N^{K-1}$. To this end, we view each $N^{K-1}$ symbols as a group and proceed until the number of symbols that remain is smaller than $N^{K-1}$,
\begin{eqnarray}
L = G_1 N^{K-1} + L_1
\end{eqnarray}
where $G_1 = \lfloor \frac{L}{N^{K-1}} \rfloor$ and $0 \leq L_1 \leq N^{K-1} - 1$. If $L_1 = 0$, we are done. Otherwise,
for the $L_1$  symbols that remain, we fully use the scheme in Theorem \ref{lemma:bia} with $N$ databases and message size $N-1$. We view each $N-1$ symbols as a group and proceed until the number of symbols left is smaller than $N-1$,
\begin{eqnarray}
L_1 = G_2 (N-1) + L_2
\end{eqnarray}
where $G_2 = \lfloor \frac{L_1}{N-1} \rfloor$ and $0 \leq L_2 \leq N - 2$. If $L_2 = 0$, we are done. Otherwise, for the $L_2 \geq 1$  symbols that are left, we use the scheme in Theorem \ref{lemma:bia}  with $L_2+1$ databases (say, the first $L_2+1 \leq N-1$ databases) and message size $L_2$. Therefore the message size and the achievable download cost are
\begin{eqnarray}
L &=& 
G_1 N^{K-1} + G_2 (N-1) + L_2 
\\
D &=& 
\left\{
\begin{array}{lc}
G_1 N^{K-1}/C + G_2 N  & \mbox{if}~ L_2 = 0 \\
G_1 N^{K-1}/C + G_2 N + L_2 + 1 &\mbox{otherwise} \\
\end{array}
\right.
\label{d}
\end{eqnarray}
This completes the description of our achievable scheme. 
\subsection{Proof that the Scheme is Correct and Private}
Since we construct our PIR scheme as a concatenation of multiple PIR schemes, let us present the following theorem to show that such a concatenation yields a PIR scheme that is correct and private. 
\begin{theorem}\label{thm:sharing}
For PIR with $N \in \mathbb{N}$ databases, each storing all $K \in \mathbb{N}$ messages, each message comprised of $L \in \mathbb{N}$ symbols from $M$-ary alphabet, $M \in \mathbb{N}/\{1\}$, where the downloads are comprised of symbols from the same $M$-ary alphabet, if there are $J \in \mathbb{N}$ schemes with message length $L_j, j \in [1:J]$ and download cost $D_j, j\in [1:J]$, respectively, and the message lengths add up to $L$, i.e., $\sum_{j=1}^J L_j = L$, then there exist a PIR scheme with message length $L$ and download cost $D = \sum_{j=1}^J D_j$.
\end{theorem}
{\it Proof:}
The scheme is based on dividing the $L$ message symbols to $J$ groups so that the $j$-th group is comprised of $L_j$ symbols per message. Then we use the given scheme with message length $L_j$ for the $j$-th group, so that the download cost achieved is $D_j$ symbols. Specifically, the queries for each group are generated independently, given  the same  desired message index. Combining the download cost for all $J$ groups, we achieve the desired download cost. We are left to prove that this symbol sharing scheme produces a correct and private PIR scheme.

Correctness is easy to see as the scheme for each group is correct. Privacy is proved as follows. Consider any database $n, n \in [1:N]$ and any desired message index $\theta, \theta \in [1:K]$. Denote the query of the scheme for the $j$-th group as $Q_n^{[\theta]}(j)$. Since the scheme for the $j$-th group is private, we have that $Q_n^{[\theta]}(j)\sim Q_n^{[\theta']}(j)$, for all $\theta,\theta'\in[1:K]$ and $\forall j\in[1:J]$. Now since for any $\theta$, the queries for each group are generated independently, their joint probability distribution function is the product of the marginal probability distribution functions, i.e.,
\begin{align}
\Pr(Q_n^{[\theta]}(1), Q_n^{[\theta]}(2), \cdots, Q_n^{[\theta]}(J)) &= \Pr(Q_n^{[\theta]}(1)) \times \Pr(Q_n^{[\theta]}(2)) \times \cdots \times \Pr(Q_n^{[\theta]}(J) ) \\
&= \Pr(Q_n^{[\theta']}(1)) \times \Pr(Q_n^{[\theta']}(2)) \times \cdots \times \Pr(Q_n^{[\theta']}(J))
\end{align}
for all $\theta,\theta'\in[1:K]$. 
Therefore the overall query for all groups is identically distributed regardless of the index of the desired message $\theta$, and the symbol sharing scheme is private (\ref{privacy}). 
\hfill\QED

\subsection{Proof that the Achieved Download Cost $D = \lceil \frac{L}{C} \rceil$}
We next show that the achievable download cost in (\ref{d}) satisfies $D \in [\frac{L}{C} , \frac{L}{C}+1)$ so that $D = \lceil \frac{L}{C} \rceil$.  Note that in the converse proof, we have already shown that for all PIR schemes, $D \geq \frac{L}{C} $ holds. So we only need to prove that $D < \frac{L}{C} + 1$. Here  we have two cases.

%
%

\noindent {\bf Case 1: $L_2 = 0$}. 
We have 
\begin{eqnarray}
D < \frac{L}{C}  + 1~\Leftrightarrow~ G_1N^{K-1}/C+G_2N&<&(G_1N^{K-1}+G_2(N-1))/C+1\\
\Leftrightarrow G_2 N &<& G_2(N-1)/C + 1 \label{eq:ww}
\end{eqnarray}
When $N = 1$, we have $G_2 = 0$ so that (\ref{eq:ww}) holds. When $N \geq 2$, plugging in $C= \frac{1-1/N}{1-(1/N)^K}=N^{K-1}\left(\frac{N-1}{N^K-1}\right)$, we have
\begin{eqnarray}
G_2N&<&G_2\left(\frac{N^K-1}{N^{K-1}}\right)+1\\
\Leftrightarrow G_2 &<& N^{K - 1}
\end{eqnarray}
which holds because $G_2 = \lfloor \frac{L_1}{N-1} \rfloor \leq L_1 \leq N^{K-1} - 1 < N^{K -1}$.  

\noindent {\bf Case 2: $L_2 \geq 1$}. 
Note that when $L_2 \geq 1$, we have $N \geq 2$ such that 
$C = \frac{1-1/N}{1-(1/N)^K}$. 
As a result,
\begin{eqnarray}
D < \frac{L}{C}  + 1~\Leftrightarrow~ G_1N^{K-1}/C+G_2N+L_2+1&<&(G_1N^{K-1}+G_2(N-1)+L_2)/C+1\\
\Leftrightarrow G_2N+L_2&<&(G_2(N-1)+L_2)/C\\
\Leftrightarrow G_2N+L_2&<&(G_2(N-1)+L_2)\left(\frac{N^K-1}{(N-1)N^{K-1}}\right)\\
\Leftrightarrow \frac{G_2}{N^{K-1}}&<&L_2\left(\frac{N^{K-1}-1}{(N-1)N^{K-1}}\right)\\
\Leftrightarrow  G_2(N-1)&<&  L_2 (N^{K-1} - 1)
\end{eqnarray}
which is proved as follows
\begin{eqnarray}
L_2 (N^{K-1} - 1) \geq N^{K-1} - 1 \geq L_1 =  G_2 (N-1) + L_2 > G_2 (N-1)
\end{eqnarray}
\noindent Thus the proof is complete.

\subsection{Proof of Theorem \ref{lemma:bia}}\label{sec:biaproof}
We now present the scheme with download cost $D = N$ and message length $L = N-1$. Consider 
\begin{eqnarray}
W_k = \Big(W_{k}(1), W_k(2), \cdots, W_{k}(N-1)\Big),\forall k \in [1:K]
\end{eqnarray}
where each $W_{k}(i), i \in [1:N-1]$ is an $M$-ary symbol. 

The queries are specified as follows.
To retrieve $W_{\theta}$ privately, the user first generates a random  vector of length $(N-1)K$, $[h_{1}(1), \cdots, h_{1}(N-1), \cdots, h_{\theta}(1), \cdots, h_\theta(N-1),\cdots, h_{K}(N-1)]$, where each element is uniformly distributed over $\{0,1\}$. Then the queries are set as follows.
\begin{eqnarray*}
Q_1^{[\theta]} &=& [h_{1}(1), \cdots, h_{\theta}(1), \cdots, h_{\theta}(N-1), \cdots, h_{K}(N-1)] \\
Q_2^{[\theta]} &=& [h_{1}(1), \cdots, h_{\theta}(1)+1, \cdots, h_{\theta}(N-1), \cdots, h_{K}(N-1)] \\
&\cdots&\\
Q_N^{[\theta]} &=& [h_{1}(1), \cdots, h_{\theta}(1), \cdots, h_{\theta}(N-1) + 1, \cdots, h_{K}(N-1)]
\end{eqnarray*}
The answer from each database is a linear combination of the message symbols, where the combining coefficients are given by the query received by that database. 
\begin{eqnarray*}
A_1^{[\theta]} &=& \sum_{k=1}^K \sum_{i=1}^{N-1} h_{k}(i) W_{k}(i) \\
A_2^{[\theta]} &=& \sum_{k=1}^K \sum_{i=1}^{N-1} h_{k}(i) W_{k}(i)  + W_{\theta}(1) \\
&\cdots&\\
A_N^{[\theta]} &=& \sum_{k=1}^K \sum_{i=1}^{N-1} h_{k}(i) W_{k}(i) + W_{\theta}(N-1) 
\end{eqnarray*}
The user decodes $ W_{\theta}(i), i \in [1:N-1]$ by subtracting $A_1^{[\theta]}$ from $A_{i+1}^{[\theta]}$, with no error. Therefore, the PIR scheme is correct. 

Privacy is guaranteed because each query is independent of the desired message index $\theta$. This is because regardless of the desired message index $\theta$, each query $Q_n^{[\theta]}, \forall n$ is individually comprised of elements that are i.i.d. uniform over $\{0,1\}$. 

Each answer is comprised of $1$ symbol, so the download cost achieved is $D = N$ symbols. The proof is complete. 

\section{Proof of Theorem \ref{thm:gen}}\label{proof:thmgen}
\subsection{Converse}
First let us prove the converse. As in the converse proof of Theorem \ref{thm:download}, the PIR capacity  \cite{Sun_Jafar_PIR} provides a general upper bound on rate, and therefore a general lower bound on download cost for any given message length, which holds regardless of the choice of alphabet used to represent the messages and download symbols.  For message length $L$ and download cost $D$, the rate is $\frac{L\log_2(M)}{D\log_2(M')}$ which cannot exceed capacity. Therefore we automatically have the lower bound on download cost as $D\geq \frac{L\log_2(M)}{C\log_2(M')}$, and because $D\in\mathbb{N}$, we must have  
\begin{eqnarray}
D\geq \left\lceil\frac{L\log_2(M)}{C\log_2(M')}\right\rceil \label{eq:generalupper}
\end{eqnarray}
\subsection{Achievability}
For the proof of achievability, let us  construct a simple (sub-optimal) PIR scheme whose download cost is nonetheless guaranteed to be within $2$ $M'$-ary symbols of the lower bound.  The scheme is described as follows. 

Let us map the messages from $M$-ary alphabet to $M'$-ary alphabet. Each message is comprised of $L$ symbols that are from an $M$-ary message alphabet, i.e., for each message there are $M^L$ possible distinct realizations. $L'$ symbols from $M'$-ary alphabet are capable of representing $M'^{L'}$ distinct realizations. To have distinct representations for distinct message realizations, we must have  ${M'}^{L'}\geq M^L$. For this, $L' = \lceil L \log_{M'} M\rceil$ is sufficient.\footnote{The sub-optimality of the scheme becomes obvious here because, for example if $M'$ is much larger than $M$, then we could jointly code all $K$ $M$-ary messages symbols to only 1 $M'$-ary message symbol, therefore download cost of $1$ $M'$-ary symbol would be enough, whereas our  naive scheme will download at least $1/C$ symbols.}  Now the message symbols and the download symbols are from the same $M'$-ary alphabet, so that we can use the PIR scheme used to establish  achievability in Theorem \ref{thm:download} to achieve download cost $D = \lceil \frac{L'}{C} \rceil$, measured in $M'$-ary download symbols. Next let us prove that even for this simple scheme, the gap to optimality is no more than $2$ $M'$-ary symbols.

Since the $N=1$ case is trivial (optimal to fully download all messages), let us assume $N\geq 2$. Note that for $N\geq 2$ it is always true that $C\geq 1/2$, i.e., $1/C\leq 2$. Starting with the general upper bound (\ref{eq:generalupper}),
\begin{eqnarray}
\left\lceil\frac{L'}{C}\right\rceil\geq D_L&\geq& \left\lceil\frac{L\log_2(M)}{C\log_2(M')}\right\rceil\\
&=&\left\lceil\frac{L\log_{M'}(M)}{C}\right\rceil\\
&\geq&\left\lceil\frac{\lceil L\log_{M'}(M)\rceil -1}{C}\right\rceil \label{eq:strict}\\
&=&\left\lceil \frac{L'}{C}-\frac{1}{C}\right\rceil\\
&\geq&\left\lceil \frac{L'}{C}-2\right\rceil\\
&=&\left\lceil \frac{L'}{C}\right\rceil-2
\end{eqnarray}

\section{Conclusion}
Recent work has characterized the capacity, $C$ (supremum of the ratio of message size over download cost, i.e., $L/D$) of PIR when the message size $L \to \infty$. 
In this work, we have shown that for arbitrary fixed message size $L \in \mathbb{N}$, when the messages and downloads are comprised of symbols from the same arbitrary $M$-ary alphabet, the optimal download cost is $D_L = \lceil \frac{L}{C} \rceil$; and when the messages and downloads are comprised of symbols from different alphabets (messages from $M$-ary alphabet and downloads from $M'$-ary alphabet, $M \neq M'$), the optimal download cost (in $M'$-ary symbols) $D_L\in\left\{\left\lceil \frac{L'}{C}\right\rceil,\left\lceil \frac{L'}{C}\right\rceil-1,\left\lceil \frac{L'}{C}\right\rceil-2\right\}$, where $L'= \lceil L \log_{M'} M\rceil$.

An interesting feature of our PIR scheme is that it allows arbitrary $M$-ary alphabet (not restricted to finite fields). This is because the scheme downloads only direct sums modulo-$M$ of various message symbols. As the next step in this direction the extension to TPIR (PIR with $T$-privacy) may be of interest. The capacity of TPIR for unconstrained alphabet is characterized in \cite{Sun_Jafar_TPIR}, and the capacity achieving scheme presented there relies on finite field operations (multiplications) and existence of MDS codes. PIR schemes based on finite fields can be extended to arbitrary $M$-ary alphabet by decomposing $M$ into its prime factors and concatenating PIR schemes over the finite fields corresponding to the prime factors. However, the extension may be difficult when field size constraints imposed by arbitrary $M$-ary alphabet are incompatible with the MDS code requirements.
\bibliographystyle{IEEEtran}
\bibliography{Thesis}
\end{document}